# Simple physical models for the partially transparent radiative windows, comparison to the radiative coolers

E. Zhang[1], C. Caloz[2], M. Skorobogatiy[1]

1 : Engineering Physics, Polytechnique Montréal, Québec, Canada.

2: Electrical Engineering, Polytechnique Montréal, Québec, Canada.

Corresponding author: maksim.skorobogatiy@polymtl.ca

**Abstract.** In this work we solve approximately the radiative heat transfer problem in one dimension to perform a comparative analysis of the time averaged performance of the partially transparent radiative windows and radiative coolers. Our physical model includes the atmosphere, the window, and the backwall that are all in the thermal equilibrium with each other, and that can exchange energy via radiative heat transfer or convection. Moreover, we use a simplified two-state model for the optical properties of an atmosphere and a window material which assumes two distinct sets of the optical reflection/absorption/transmission parameters in the visible/near-IR versus mid-IR spectral ranges. Furthermore, we have distinguished the design goals for the partially transparent windows and radiative coolers and provided optimal choice for the material parameters to realize these goals. Thus, radiative coolers are normally non-transparent in the visible, and the main goal is to design a cooler with the temperature of its dark side as low as possible compared to that of the atmosphere. For the radiative windows, however, their surfaces are necessarily partially transparent in the visible. In the cooling mode, therefore, the main question is about the maximal visible light transmission through the window at which the temperature on the window somber side does not exceed that of the atmosphere. We believe that our simple, yet in-depth comparative analysis of the radiative windows and coolers can be useful for a large number of scientists and engineers pursuing research in these disciplines.

## 1. Introduction

Currently, the problem of energy overconsumption presents one of the major challenges for many societies [1]. In particular, energy spending for building cooling, heating and lighting accounts for ~40% of the total energy consumption in some countries [2, 3]. While several methods to reduce energy use for the cooling of buildings, automobiles, food depots and other structures are known, radiative cooling distinguishes itself by being one of the few temperature reduction methods that require no external energy sources for their operation [1, 2]. Design of radiative coolers presents an intriguing multidisciplinary problem at the intersection of physics, material science, optics and engineering.

Similarly to other cooling methods, general idea of the radiative cooling is to achieve larger energy outgoing flux than the incoming energy flux [4]. Standard passive cooling devices use large heat sinks with lower temperature than that of a cooled object to reach large outgoing energy fluxes [5]. On a planetary scale, the Earth temperature is ~300K, which is much larger than the temperature of the outer space ~3 K that acts as an infinite heat sink for the planet [4]. As the outer space is a near perfect vacuum, the main heat loss mechanism for the Earth is not a conduction driven heat transfer but a radiative one. This is due to the fact that objects with temperatures above the absolute zero radiate electromagnetic energy known as thermal radiation or black-body radiation with the spectrum depending on the object temperature and described by the Planck's law. Thus, objects with temperatures of ~300 K mainly radiate energy in the mid-IR with a broad peak in the spectral density covering the wavelength of 8-15 μm [5]. Remarkably, in the same spectral rage, the Earth's atmosphere has a transmission window (window of transparency) that allows a considerable portion of the mid-IR thermal radiation to leave unimpeded [5,6]. This represents the main mechanism by which terrestrial objects can dissipate heat into the outer space in the form of electromagnetic waves, which is the basis of radiative cooling. It is pertinent to mention at this point that the Earth cooling via radiative heat transfer to the outer space is compensated by heating of the Earth surface via partial absorption of the radiative energy of the Sun that features the highest spectral density in the visible, while extending all the way to the near-IR 0.3-2.5 μm [7].

The study of radiative cooling has a long history. It has been well known since the ancient times that using blackened surfaces (radiators) facing a clear night sky could result in sub-ambient temperatures of the radiator, which was even used to make ice [8, 9]. Temperature reduction of about 5℃ was reported by Granqvist in 1981 [10], using a low emittance window surface and night time cooling. This original work was followed by several studies of radiative cooling efficiency under different environmental conditions, such as humidity [11, 12], ambient temperature [13], and geographical location [14]. This original work was followed by studies of the day time cooling, which turned to be a much harder problem. Originally, daytime radiative cooling under direct sunlight was achieved by using a radiator that would reflect most of the sunlight, while radiating efficiently in the mid-IR spectral range [15], thus allowing the heat to escape through the atmospheric transparency window. The fact that the radiative energy of the Sun and the thermal radiation of a terrestrial object occupy different and mostly non-overlapping parts of the electromagnetic spectrum make the problem of the radiative cooler design more complicated. This is because the cooler materials have to exhibit widely different thermo-optic properties in the visible and mid-IR spectral ranges.

A large number of studies have been conducted to date on a subject of radiative cooler materials (electrochromic, photochromic windows and thermochromic windows [16-18], photonic crystals and metamaterials [19,20]), as well as structures and design optimization (reflecting vs. absorbing structures [21-23]), with experimentally demonstrated temperature reduction from 5℃ to 42℃ [15, 24, 25]. Furthermore, the question of the fundamental limit in the temperature reduction for radiative coolers was investigated in great details with a consensus that it depends strongly on the environmental conditions [12, 26]. Thus, for the night time radiative coolers the predictions range between 15-42°C [27-29], while for the daytime radiative coolers the temperature reduction is expected to be only several degrees [15].

An important issue when characterizing radiating coolers is the choice of parameters to characterize the cooler performance. While the cooler itself has a working surface, its main function is to cool the air and solid objects behind it. Therefore, while the cooler surface temperature is of importance the more practical parameter is probably the air temperature or the temperature of a solid behind the cooler. Another issue is about on-average versus instantaneous performance of a cooler. While some coolers work mostly during night-time, others are capable of the day and night operation [30]. They are known as high efficiency radiative coolers or radiative windows if the cooler surface is partially transparent. Additionally, if the window can adjust its radiative properties in response to the changing environmental conditions such as ambient temperature and daylight illumination [31-33], or if it is capable of both cooling and heating, such windows are frequently referred to as intelligent or smart window [34].

Recently, there has been a strong interest in passive and active smart windows that could operate year-long, while providing heating in the cold and cooling in the hot weathers [30,35]. Much research is focusing on developing materials that can simultaneously, while independently manage radiation across several widely spaced spectral ranges covering, for example, visible / near-IR light [14, 36, 37], or solar / mid-IR light [25].

At the same time, a concept of perfect smart window was proposed to judge the energy efficiency of the existing smart windows [38, 39]. Thus, the perfect smart window has zero absorptivity for the visible light, near infrared and mid infrared. At the same time, the window features a perfect transmittance in the visible, while the mid-IR light transmittance and reflectance should be either (zero, perfect) or (perfect, zero) depending on whether heating or cooling is required. In any case, to switch between the heating and cooling states the intelligent windows should allow large difference in its transmittance/reflectance properties between the two states, while always featuring low absorption of solar radiation [38]. Therefore, much attention was payed to improving the performances of the smart windows through materials research and structural optimization [35, 40-45].

While much work has been done on theoretical understanding of the functioning of radiative coolers, in the related field of smart windows the performance targets and optimization strategies are still less understood. As functioning of the smart windows is dominated by the multi-year-long time scale, therefore it is interesting to analyze their averaged-over-time performance rather than an instantaneous response. It is also important to acknowledge that the function of a window implies nonzero visible light transmission, therefore a tradeoff between the window performance and its esthetic function seems unavoidable.

In this paper we consider in more details partially transparent radiative coolers/heaters in the context of their application in smart windows. From the onset we distinguish the design goals for the radiative coolers from those of the partially transparent (in the visible) radiative windows. Thus, radiative coolers are normally non-transparent in the visible, and the main goal is to design a cooler with the temperature of its dark side as low as possible compared to that of the atmosphere. For the radiative windows, however, their surfaces are necessarily partially transparent in the visible. In the cooling mode, therefore, the main question is about the maximal visible light transmission through the window at which the temperature on the window somber side does not exceed that of the atmosphere. Alternatively, as a measure of the window performance one can use the temperature of the adjacent inner air region or even that of the room wall. Finally, one needs to acknowledge that in practical application, heat convection is always present and its contribution to the smart window performance has to be quantified.

Ideally, in order to model precisely the smart window performance, one has to have a realistic model of the atmospheric optical properties [46], and then solve a full radiative heat transfer problem [47] that includes atmosphere, window and enclosure where the window is installed. While certainly possible, such an overwhelming approach will most probably obscure the relatively simple physics behind the problem. Therefore, in this work we strip the radiative heat transfer model to a bare minimum and confine ourselves to one dimension, while still retaining all the key elements of a problem. Particularly, we consider the atmosphere, the window and the backwall that are all in the thermal equilibrium and that all exchange the energy via radiative heat transfer or convection. Moreover, we use a simplified two-state model for the optical properties of an atmosphere and a window material that assumes two distinct sets of the optical reflection/absorption/transmission parameters in the visible/near-IR versus mid-IR spectral ranges. We believe that our work captures all the important aspects of the radiative heat transfer in smart windows, and allows to make predictions on the choice of the windows optimal design parameters.

## 2. Model 1. Single layer model of the atmosphere.

We start with a single layer model of the atmosphere [48]. This model allows to relate together the Earth and the atmosphere average temperatures, the atmosphere average emissivity, the planet albedo (reflectivity) and the total incoming radiative flux from the Sun. The model assumes equilibrium temperature distribution, and, hence, net zero radiative heat flux through any interface parallel to the Earth. A key element of a model is a recognition that in equilibrium, material absorption coefficient equals to the material emission coefficient. This model is a necessary point of departure that allows to define a self-consistent reference (average atmospheric properties versus average Sun radiative flux) for the following models of a smart window.

Notation:
$E(\nu, T)$ - emission spectrum (by power) of a blackbody at temperature $T$ (see appendix)
$P(\nu)$ - energy flux spectrum of the incoming light from the Sun

Atmosphere -
$T_a$ - temperature of the atmosphere
$L_a$ - thickness of the atmosphere
$r_a(\nu)$ - frequency dependent reflection coefficient of the atmosphere by power (mostly due to Rayleigh scattering)
$\alpha_a(\nu)$ - material absorption coefficient of the atmosphere by power
$t_a(\nu)$ - frequency dependent transmission coefficient of the atmosphere by power:

$$t_a(\nu) \approx (1 - r_a(\nu)) \exp(-\alpha_a(\nu) L_a)$$

$\varepsilon_a(\nu)$ - emissivity of the atmosphere at frequency $\nu$ equals to its absorption:

$$\varepsilon_a(\nu) = a_a(\nu) = 1 - r_a(\nu) - t_a(\nu)$$

The Earth -
$T_g$ - temperature of the Earth ground

We assume that emissivity of the Earth at any frequency is 1 (blackbody approximation)

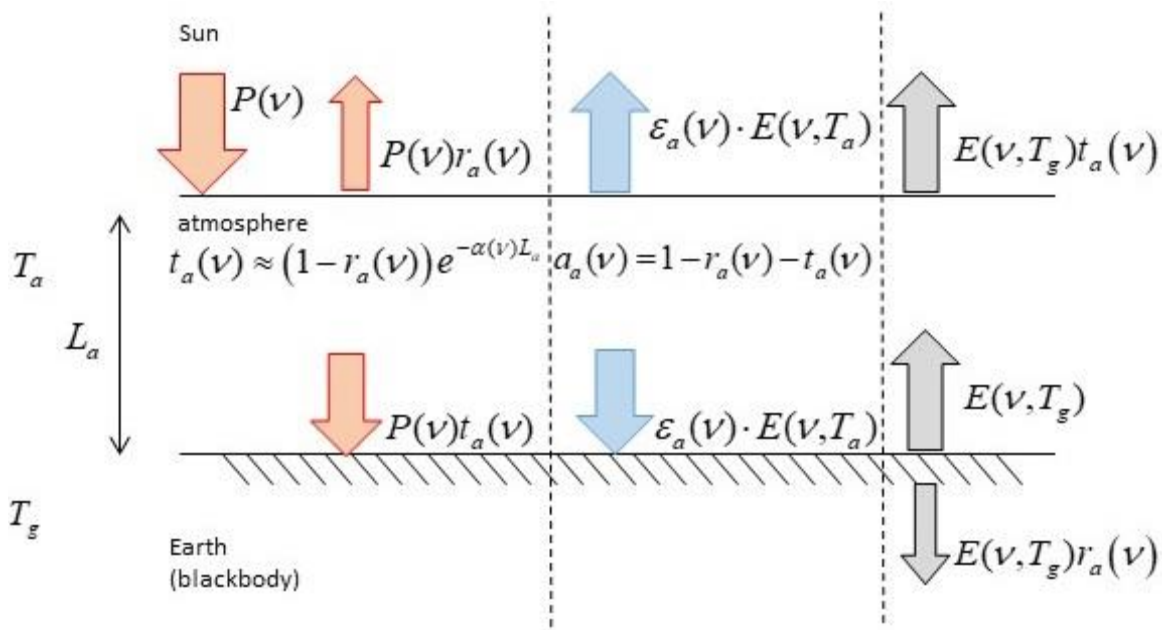

Fig. 1. Single layer model of the atmosphere.

Assuming a constant temperature in the atmosphere, we now use the energy conservation principle to write the following equations:

At the space edge of the atmosphere we write for the energy fluxes in equilibrium:

$$\int (1 - r_a(v)) \cdot P(v)\, dv = \int \varepsilon_a(v) \cdot E(v, T_a)\, dv + \int t_a(v) \cdot E(v, T_g)\, dv \quad (1)$$

At the Earth level we write for the energy fluxes in equilibrium:

$$\int t_a(v) \cdot P(v)\, dv = -\int \varepsilon_a(v) \cdot E(v, T_a)\, dv + \int (1 - r_a(v)) \cdot E(v, T_g)\, dv \quad (2)$$

Alternatively, by subtracting (2) from (1) we also get the balance between the absorbed and irradiated energy in the atmosphere:

$$2\int \varepsilon_a(v) \cdot E(v, T_a)\, dv = \int a_a(v) \cdot P(v)\, dv + \int a_a(v) \cdot E(v, T_g)\, dv \quad (3)$$

Now, we introduce several approximations for the values of the various parameters involved in the model (1)-(3) using physical properties of the atmosphere.

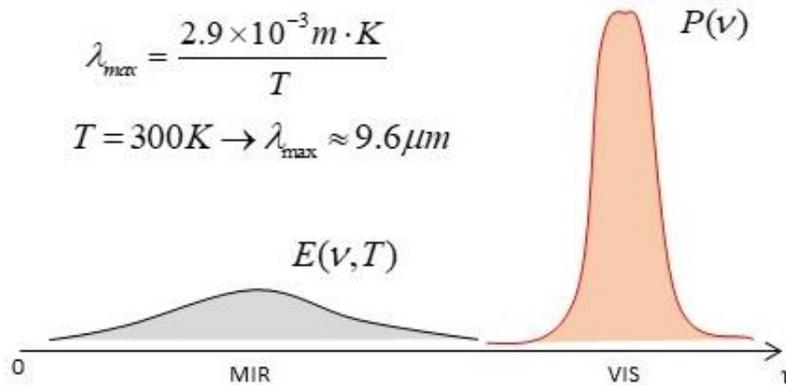

Fig. 2. Power of the incoming light is mostly concentrated in the visible spectral range, while power of the irradiated light by the atmosphere and Earth are mostly in the Mid-IR spectral range.

First, we note that power of the incoming light from the Sun is mostly concentrated in the visible spectral range, while power of the irradiated light by the atmosphere and Earth are mostly in the mid-IR spectral range, with little overlap between the two (see Fig. 2). Therefore, instead of using frequency-variable transmission and reflection properties of the atmosphere, we rather assume a step-like spectral behavior of its optical properties. Particularly, we assume that in the visible and in the mid-IR spectral ranges, the optical properties of the atmosphere are distinct from each other and frequency invariable as shown in Fig. 3.

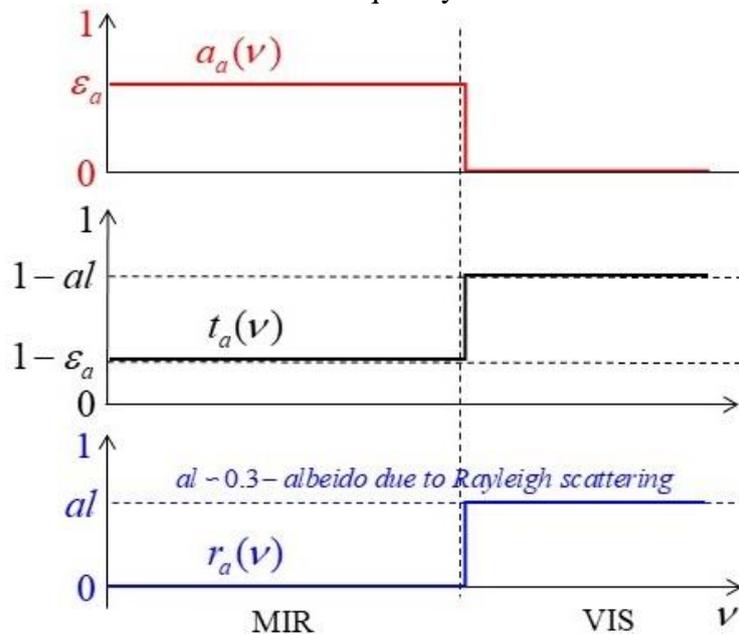

Fig. 3. Step-like model for the frequency dependent optical properties of the atmosphere (absorption, transmission and reflection).

With thus defined properties of the atmosphere we get:

$$\int t_a(\nu) \cdot P(\nu) d\nu = (1-a) \int P(\nu) d\nu = (1-al)\bar{P}$$

$$\int (1-r_a(\nu)) \cdot P(\nu) d\nu = (1-al)l\bar{P}$$

$$\int (1-r_a(\nu)) \cdot E(\nu, T_g) d\nu = \sigma T_g^4$$

$$\int t_a(\nu) \cdot E(\nu, T_g) d\nu = (1-\varepsilon_a)\sigma T_g^4 \quad (4)$$

$$\int \varepsilon_a(\nu) \cdot E(\nu, T_a) d\nu = \varepsilon_a \sigma T_a^4$$

$$\int a_a(\nu) \cdot P(\nu) d\nu = 0$$

$$\int a_a(\nu) \cdot E(\nu, T_g) d\nu = \varepsilon_a \sigma T_g^4$$

while equations (1), (2) and (3) (only two of the three are independent) become:

$$\begin{cases} (1-al)\bar{P} = \varepsilon_a \sigma T_a^4 + (1-\varepsilon_a)\sigma T_g^4 \\ (1-al)\bar{P} = -\varepsilon_a \sigma T_a^4 + \sigma T_g^4 \\ 2 \cdot T_a^4 = T_g^4 \end{cases} \quad (5)$$

from which is follows that:

$$\sigma T_a^4 = \frac{(1-al)}{(2-\varepsilon_a)}\bar{P} \quad (6)$$

$$T_g = 2^{1/4} \cdot T_a$$

In order to reproduce with this model the average Earth temperature $T_g = 288.2$, while using for the planet albedo $al = 0.3$, and the average radiative flux from the Sun incident onto the planet $\bar{P} = 342 W \cdot m^{-2}$, one requires to choose for the atmosphere emissivity $\varepsilon_a = 0.78$. This also results in a somewhat low atmospheric temperature $T_a = 242.11 K$ ($-30°C$), which is a well know deficiency of a single layer atmospheric model. Note that the quoted value for the average incoming solar radiation takes into account the angle at which the rays strike and that at any one moment half the planet does not receive any solar radiation. It therefore measures only one-fourth of the solar constant, which is an averaged over the year energy flux incident on the Earth as measured by an orbiting satellite.

### 3. Model 2. Single layer optically symmetric window, no convection.

We now consider a single layer optically symmetric window placed in the path of a sunlight. The window is assumed to be optically thick (no interference effects). We also assume that the temperature across the window is constant (see discussion in Appendix A), and, therefore, the thermal radiation from the window is the same in both directions. Behind the window we place a perfect absorber that we refer to as a wall. The wall is assumed to be thermally and radiationally separated from the ground. We, furthermore, assume that there is vacuum between the window and the wall, so we neglect the convection and conduction heat transfer in this region. Finally, the Earth atmosphere is characterized by the average temperature and permittivity defined in the previous section.

Notation:

$P_g(\nu) = t_a(\nu) \cdot P(\nu)$ - spectrum of the incoming radiation from the sun at the ground level by power

$\bar{\bar{P}} = (1-al)\bar{P}$ - average power of the sun at the planet (window) surface

Window -

$T_w$ - temperature of the window

$r_w(\nu)$ - frequency dependent reflection coefficient of the windows by power

$t_w(\nu)$ - frequency dependent transmission coefficient of the windows by power

$a_w(\nu)$ - absorption of the atmosphere at frequency $\nu$ equals to its emissivity:

$$\varepsilon_w(\nu) = a_w(\nu) = 1 - r_w(\nu) - t_w(\nu)$$

Wall -

$T_s$ - temperature of the wall surface behind the window which is considered to be a blackbody.

Spectral dependence of the window material parameters are assumed to be step-like, with two distinct sets of frequency independent parameters defining window properties in the mid-IR and visible ranges as shown in Fig. 4.

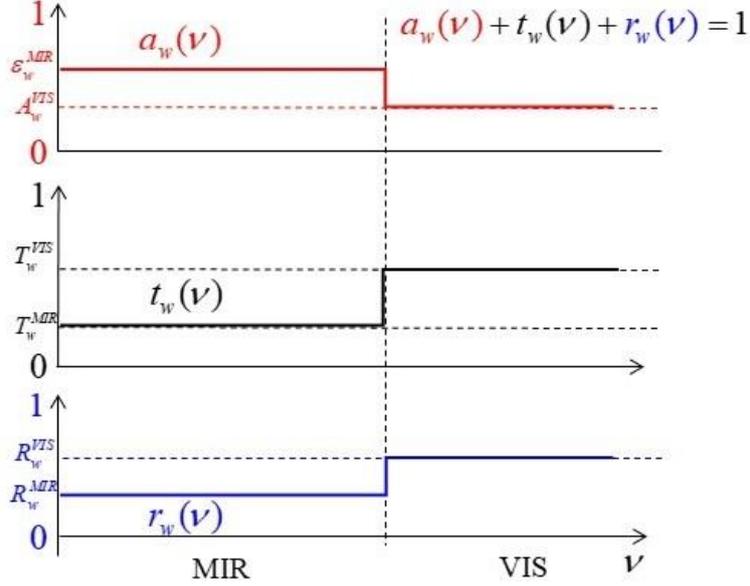

Fig. 4. Step-like model for the frequency dependent optical properties of the window (absorption, transmission and reflection).

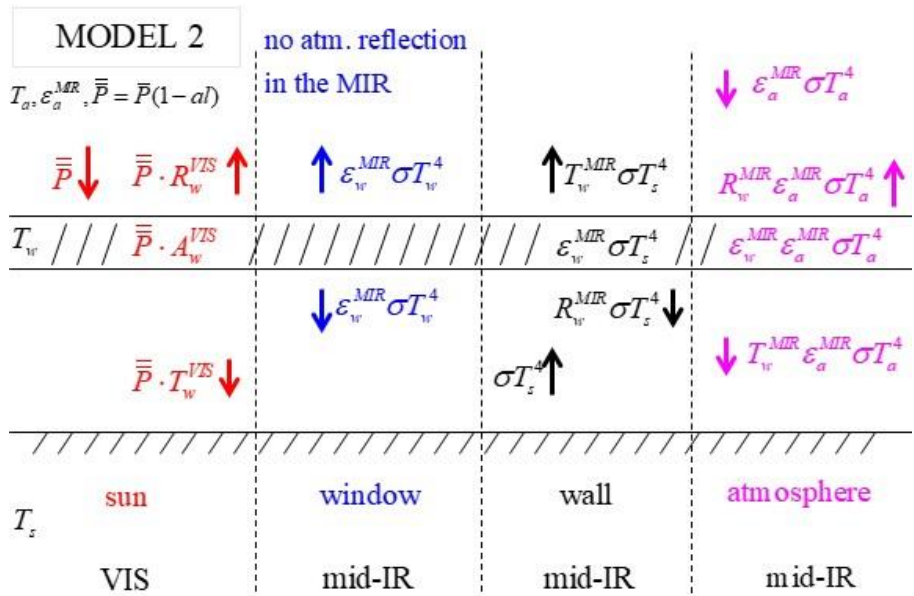

Fig. 5. Model of the optically symmetric single layer window (no convection/conduction between the wall and the window).

We now use the energy conservation principle to write the following equations:

At the window / atmosphere interface we write for the energy fluxes in equilibrium:
$$\int (1 - r_w(\nu)) \cdot P_g(\nu)\, d\nu = \int \varepsilon_w(\nu) \cdot E(\nu, T_w)\, d\nu + \int t_w(\nu) \cdot E(\nu, T_s)\, d\nu - \int (1 - r_w(\nu)) \cdot \varepsilon_a(\nu) \cdot E(\nu, T_a)\, d\nu \quad (7)$$

At the wall (blackbody) level we write for the energy fluxes in equilibrium:
$$\int t_w(\nu) \cdot P_g(\nu)\, d\nu = -\int \varepsilon_w(\nu) \cdot E(\nu, T_w)\, d\nu + \int (1 - r_w(\nu)) \cdot E(\nu, T_s)\, d\nu - \int t_w(\nu) \cdot \varepsilon_a(\nu) \cdot E(\nu, T_a)\, d\nu \quad (8)$$

Alternatively, by subtracting (8) from (7) we also get the balance between the absorbed and irradiated energy in the window:

$$2\int \varepsilon_w(\nu) \cdot E(\nu, T_w) \, d\nu = \int a_w(\nu) \cdot P(\nu) \, d\nu + \int a_w(\nu) \cdot E(\nu, T_s) \, d\nu + \int a_w(\nu) \cdot \varepsilon_a(\nu) \cdot E(\nu, T_a) \, d\nu \quad (9)$$

These equations can be further simplified when using a step-like model for the material optical properties (only two of the three following equation are independent):

$$\begin{cases} (1 - R_w^{VIS}) \cdot \overline{\overline{P}} = \varepsilon_w^{MIR} \cdot \sigma T_w^4 + T_w^{MIR} \cdot \sigma T_s^4 - (1 - R_w^{MIR}) \cdot \varepsilon_a \sigma T_a^4 \\ T_w^{VIS} \cdot \overline{\overline{P}} = -\varepsilon_w^{MIR} \cdot \sigma T_w^4 + (1 - R_w^{MIR}) \cdot \sigma T_s^4 - T_w^{MIR} \cdot \varepsilon_a \sigma T_a^4 \\ 2\varepsilon_w^{MIR} \cdot \sigma T_w^4 = A_w^{VIS} \cdot \overline{\overline{P}} + \varepsilon_w^{MIR} \cdot \sigma T_s^4 + \varepsilon_w^{MIR} \cdot \varepsilon_a \sigma T_a^4 \end{cases} \quad (10)$$

Now, using (10), as well as results of the model 1 that relates average atmospheric temperature to the average power flux arriving at the planet surface, we can arrive to the following expressions for the temperatures of the window and the wall:

$$\frac{T_w^4}{T_a^4} = \varepsilon_a + (2 - \varepsilon_a) \frac{T_w^{VIS} + A_w^{VIS} \cdot \left(1 + \frac{T_w^{MIR}}{\varepsilon_w^{MIR}}\right)}{2T_w^{MIR} + \varepsilon_w^{MIR}}$$

$$\frac{T_s^4}{T_a^4} = \varepsilon_a + (2 - \varepsilon_a) \frac{2T_w^{VIS} + A_w^{VIS}}{2T_w^{MIR} + \varepsilon_w^{MIR}}$$
(11)

### 3.1 Analysis of the Model 2

1) A particular solutions of Model 1 (Eq. 11) is $T_w = T_s = T_a$ when choosing $T_w^{VIS} = \gamma \cdot T_w^{MIR}$, $A_w^{VIS} = \gamma \cdot \varepsilon_w^{MIR}$, for any values of $T_w^{MIR}, \varepsilon_w^{MIR}$, where $\gamma = \frac{1 - \varepsilon_a}{2 - \varepsilon_a} = 0.18$. Particularly, in the case of highly transmissive windows in the mid-IR $T_w^{MIR} \to 1$, which simultaneously feature low absorption in the visible $A_w^{VIS} = \gamma \cdot \varepsilon_w^{MIR} = \gamma \cdot (1 - T_w^{MIR} - R_w^{MIR}) \to 0$, window transmission of the visible light can be as high as $T_w^{VIS} = \gamma = 18\%$, while both the room and the wall temperature will not exceed that of the atmosphere.

2) If the window is an efficient absorber or reflector, so that there is no transmission through the window both in the visible and in the mid-IR $T_w^{VIS} = T_w^{MIR} = 0$, then:

$$\frac{T_s^4}{T_a^4} = \frac{T_w^4}{T_a^4} = \varepsilon_a + (2 - \varepsilon_a) \frac{A_w^{VIS}}{\varepsilon_w^{MIR}} \quad (12)$$

Particularly, if the window absorption in the visible is much smaller that the window absorption in the mid-IR, then the window is an efficient radiative cooler with the window and wall temperatures smaller than that of an atmosphere:

$$A_w^{VIS} \ll \varepsilon_w^{MIR} \Rightarrow T_s = T_w \to \varepsilon_a^{1/4} \cdot T_a \approx 0.94 \cdot T_a$$

Alternatively, if the window is a black body or a "balanced" vis/mid-IR absorber, then the window and wall temperature will be those of a ground:

$$A_w^{VIS} = \varepsilon_w^{MIR} \Rightarrow T_s = T_w = 2^{1/4} T_a = T_g$$

Finally, if the window is a strong absorber in the visible, but a week absorber in the mid-IR, it becomes an efficient radiative heater:

$$A_w^{VIS} > \varepsilon_w^{MIR} \Rightarrow T_s = T_w > T_g$$

3) What is the smallest wall temperature $T_s$ possible? From (11) it follows that:

$$\min(T_s) = \varepsilon_a^{1/4} \cdot T_a \approx 0.94 \cdot T_a$$

when $\dfrac{2T_w^{VIS} + A_w^{VIS}}{2T_w^{MIR} + \varepsilon_w^{MIR}} = 0$ (13)

$\Rightarrow T_w^{VIS} \to 0 \, ; \, A_w^{VIS} \to 0 \Rightarrow R_w^{VIS} \to 1$

we must also require $2T_w^{MIR} + \varepsilon_w^{MIR} = 1 + T_w^{MIR} - R_w^{MIR} \neq 0 \Rightarrow R_w^{MIR} \neq 1$

This is the case of an almost completely reflective window in the visible, which at the same time has a non-perfect reflectivity in the mid-IR.

4) What is the smallest window temperature $T_w$ possible? From (11) it follows that:

$$\min(T_w) = \varepsilon_a^{1/4} \cdot T_a \approx 0.94 \cdot T_a$$

when $\dfrac{T_w^{VIS} + A_w^{VIS} \cdot \left(1 + \dfrac{T_w^{MIR}}{\varepsilon_w^{MIR}}\right)}{2T_w^{MIR} + \varepsilon_w^{MIR}} = 0 \Rightarrow T_w^{VIS} \to 0 \, ; \, A_w^{VIS} \to 0 \, ; \, \dfrac{A_w^{VIS}}{\varepsilon_w^{MIR}} \to 0$ (14)

$\Rightarrow R_w^{VIS} \to 1 \, ; \, A_w^{VIS} \ll \varepsilon_w^{MIR}$

we must also require $2T_w^{MIR} + \varepsilon_w^{MIR} = 1 + T_w^{MIR} - R_w^{MIR} \neq 0 \Rightarrow R_w^{MIR} \neq 1$

This is the case of an almost completely reflective window in the visible, which at the same time has a non-perfect reflectivity in the mid-IR. Additionally we have to require that the window absorption in the visible should be much smaller than the window absorption in the mid-IR.

5) What is the maximal transmission of the visible light through the window $T_w^{VIS}$, so that the wall temperature does not exceed the atmospheric temperature $T_s < T_a$. From (11) it follows that:

$$\dfrac{T_s^4}{T_a^4} = \varepsilon_a + (2 - \varepsilon_a) \dfrac{2T_w^{VIS} + A_w^{VIS}}{2T_w^{MIR} + \varepsilon_w^{MIR}} < 1$$

$$T_w^{VIS} < \dfrac{1}{2} \dfrac{1 - \varepsilon_a}{2 - \varepsilon_a} \left(1 + T_w^{MIR} - R_w^{MIR}\right) - \dfrac{A_w^{VIS}}{2}$$ (15)

$\max(T_w^{VIS}) = \gamma$ when $T_w^{MIR} = 1$ ($\varepsilon_w^{MIR} = R_w^{MIR} = 0$) ; $A_w^{VIS} = 0$

where $\gamma = \dfrac{1 - \varepsilon_a}{2 - \varepsilon_a} = 0.18$

In order to achieve maximal transmission of the visible light, while still having the wall temperature below that of the atmosphere, we have to demand that window absorption in the visible is zero $A_w^{VIS} \to 0$, while window transmission in the mid-IR is almost perfect $T_w^{MIR} \to 1$.

6) What is the maximal transmission of the visible light through the window $T_w^{VIS}$, so that the window temperature does not exceed the atmospheric temperature $T_w < T_a$. From (11) it follows that:

$$\dfrac{T_w^4}{T_a^4} = \varepsilon_a + (2 - \varepsilon_a) \dfrac{T_w^{VIS} + A_w^{VIS} \cdot \left(1 + \dfrac{T_w^{MIR}}{\varepsilon_w^{MIR}}\right)}{2T_w^{MIR} + \varepsilon_w^{MIR}} < 1$$

$$T_w^{VIS} < \dfrac{1 - \varepsilon_a}{2 - \varepsilon_a} \left(1 + T_w^{MIR} - R_w^{MIR}\right) - A_w^{VIS} \cdot \left(1 + \dfrac{T_w^{MIR}}{\varepsilon_w^{MIR}}\right)$$ (16)

$\max(T_w^{VIS}) = 2\gamma = 0.36$ when $T_w^{MIR} \to 1$ ($\varepsilon_w^{MIR}, R_w^{MIR} \to 0$) ; $A_w^{VIS} \ll \varepsilon_w^{MIR} \to 0$

In order to achieve maximal transmission of the visible light, while still having the window temperature below that of the atmosphere, we have to demand that window transmission in the mid-IR is almost perfect $T_w^{MIR} \to 1$, while window absorption in the visible is much smaller than that in the mid-IR, and at the same time approaching zero $A_w^{VIS} \to 0$.

7) Tradeoff between window transmission and absorption in the visible can be summarized geometrically in Fig. 6. As follows from inequalities (15) and (16), shaded regions below the corresponding lines represent the range of values of $T_w^{VIS}, A_w^{VIS}$ for which either the wall temperature or the window temperature are smaller than the atmospheric temperature. Intersection of these two regions represent the range of values of $T_w^{VIS}, A_w^{VIS}$ for which both the wall temperature and the window temperature are smaller than the atmospheric one. Both the window and the wall temperatures attain that of the atmospheric one when $T_w^{VIS} = \gamma \cdot T_w^{MIR}$, $A_w^{VIS} = \gamma \cdot \varepsilon_w^{MIR}$. As clear from the graph, when $T_w^{VIS} \leq \gamma \cdot T_w^{MIR}$ and $A_w^{VIS} \leq \gamma \cdot \varepsilon_w^{MIR}$, it is guaranteed that both the window and the wall temperatures will be smaller than the atmospheric temperature.

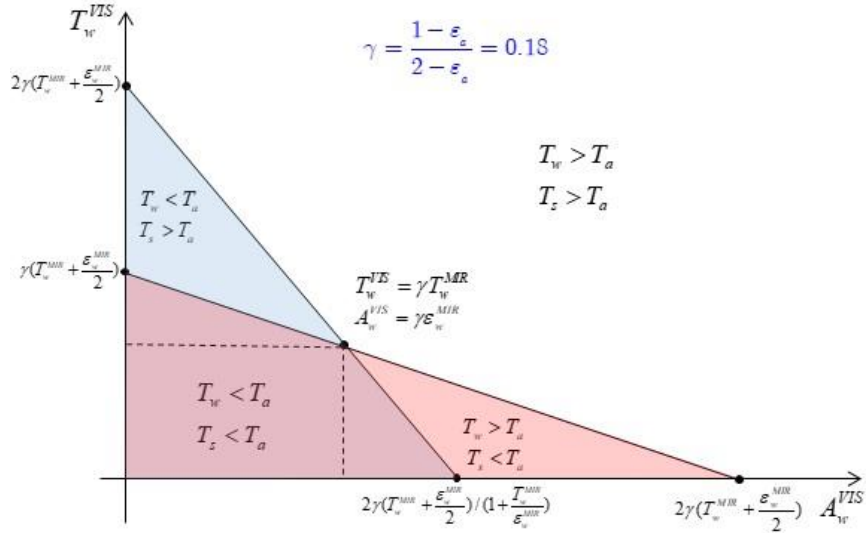

Fig. 6. Tradeoff between window transmission and absorption in the visible (no convection). A region of parameter space that is situated below the two lines define window operation regime for which both the wall and the window temperatures are smaller than that of the atmosphere.

## 4. MODEL 3. Single layer optically symmetric window, with convection.

In this model we add convection heat transfer between window and air that surrounds it, while ignoring heat conduction through the air. The model is virtually identical to the Model 2, with only two new parameters, which are the room temperature (space between the window and the wall), and the heat transfer coefficient. Value of the heat transfer coefficient is in general temperature dependent and can be significantly different for horizontal or vertical surfaces. In Appendix B, for completeness, we present several well-known models for the heat transfer coefficients at the planar interfaces between gas and solid.

Notation:

Room –

$T_r$ - temperature of the air-filled space between window and a blackbody wall.

$h$ - temperature dependent heat transfer coefficient that defines convection at the solid/air surfaces

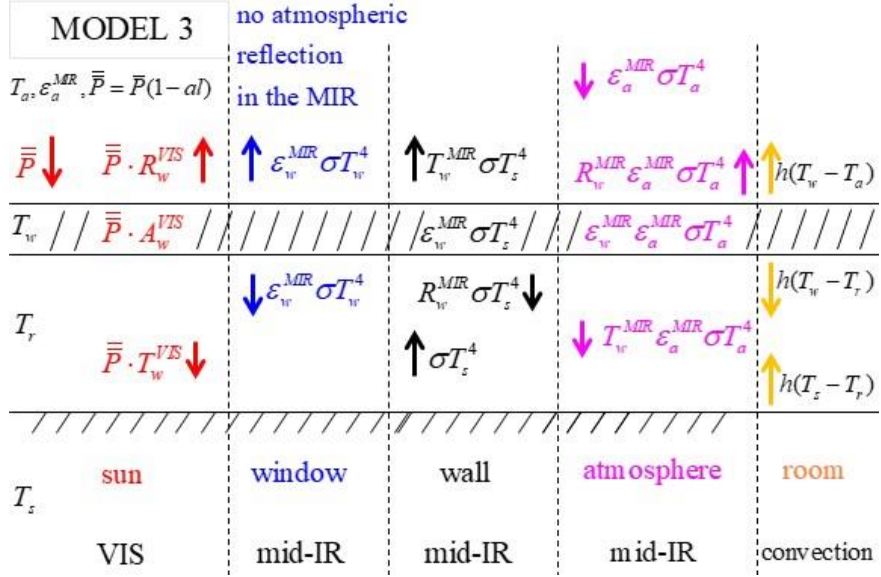

Fig. 7. Model of the optically symmetric single layer window (with convection).

We now use the energy conservation principle to write the following equations:

At the window / atmosphere interface we write for the energy fluxes in equilibrium:
$$\int (1-r_w(\nu))\cdot P_g(\nu)\, d\nu = \int \varepsilon_w(\nu)\cdot E(\nu,T_w)\, d\nu + \int t_w(\nu)\cdot E(\nu,T_s)\, d\nu - \int (1-r_w(\nu))\cdot \varepsilon_a(\nu)\cdot E(\nu,T_a)\, d\nu + h(T_w - T_a) \quad (17)$$

At the window / room interface we write for the energy fluxes in equilibrium:
$$\int t_w(\nu)\cdot P_g(\nu)\, d\nu = -\int \varepsilon_w(\nu)\cdot E(\nu,T_w)\, d\nu + \int (1-r_w(\nu))\cdot E(\nu,T_s)\, d\nu - \int t_w(\nu)\cdot \varepsilon_a(\nu)\cdot E(\nu,T_a)\, d\nu - h(T_w - T_r) \quad (18)$$

By subtracting (18) from (17) we also get the balance between the absorbed and irradiated energy in the window, as well as heat transferred into the window due to convection:
$$2\int \varepsilon_w(\nu)\cdot E(\nu,T_w)\, d\nu = \int \alpha_w(\nu)\cdot P(\nu)\, d\nu + \int \alpha_w(\nu)\cdot E(\nu,T_s)\, d\nu + \int \alpha_w(\nu)\cdot \varepsilon_a(\nu)\cdot E(\nu,T_a)\, d\nu + h(2T_w - T_a - T_r) \quad (19)$$

Finally, in thermal equilibrium, two convection flows inside of the room have to cancel each other:
$$h(T_s - T_r) + h(T_w - T_r) = 0 \Rightarrow T_r = \frac{T_s + T_w}{2} \quad (19)$$

Assuming step-like spectral properties of the window material optical properties in (17-19) (see Fig. 4) we can then write the following system of equations (only 2 of them are independent):
$$\begin{cases} (1-R_w^{VIS})\cdot \bar{\bar{P}} = \varepsilon_w^{MIR}\cdot \sigma T_w^4 + T_w^{MIR}\cdot \sigma T_s^4 - (1-R_w^{MIR})\cdot \varepsilon_a \sigma T_a^4 + h(T_w - T_a) \\ T_w^{VIS}\cdot \bar{\bar{P}} = -\varepsilon_w^{MIR}\cdot \sigma T_w^4 + (1-R_w^{MIR})\cdot \sigma T_s^4 - T_w^{MIR}\cdot \varepsilon_a \sigma T_a^4 - h(T_w - T_r) \\ 2\varepsilon_w^{MIR}\cdot \sigma T_w^4 = A_w^{VIS}\cdot \bar{\bar{P}} + \varepsilon_w^{MIR}\cdot \sigma T_s^4 + \varepsilon_w^{MIR}\cdot \varepsilon_a \sigma T_a^4 + h(T_w - T_a) + h(T_w - T_r) \end{cases} \quad (21)$$

Now, using results of the Model 1 that relates atmospheric temperature to the average light power arriving at the planet surface, we can arrive to the following coupled nonlinear equations for the temperatures of the window and the wall:

$$\frac{T_w^4}{T_a^4} + \frac{h}{\sigma T_a^3 (2T_w^{MIR} + \varepsilon_w^{MIR})}\left(\frac{T_w - T_a}{T_a} + \frac{3T_w - T_s - 2T_a}{2T_a}\frac{T_w^{MIR}}{\varepsilon_w^{MIR}}\right) = \varepsilon_a + (2-\varepsilon_a)\frac{T_w^{VIS} + A_w^{VIS}\cdot\left(1 + \frac{T_w^{MIR}}{\varepsilon_w^{MIR}}\right)}{2T_w^{MIR} + \varepsilon_w^{MIR}}$$
(22)

$$\frac{T_s^4}{T_a^4} + \frac{h}{\sigma T_a^3 (2T_w^{MIR} + \varepsilon_w^{MIR})}\left(\frac{T_w + T_s - 2T_a}{2T_a}\right) = \varepsilon_a + (2-\varepsilon_a)\frac{2T_w^{VIS} + A_w^{VIS}}{2T_w^{MIR} + \varepsilon_w^{MIR}}$$

These equations can be solved analytically by linearizing the nonlinear terms assuming that the wall and window temperatures are close to the atmospheric temperature $\frac{T_{s,w}^4}{T_a^4} \approx 1 + 4\frac{\delta T_{s,w}}{T_a}$, where $\delta T_{s,w} = T_{s,w} - T_a$.

In this case, a system of nonlinear equations (22) becomes linear:

$$4\frac{\delta T_w}{T_a} + \underbrace{\frac{h}{2\sigma T_a^3 (2T_w^{MIR} + \varepsilon_w^{MIR})}}_{\chi} \left( \frac{\delta T_w}{T_a}\left(2 + 3\frac{T_w^{MIR}}{\varepsilon_w^{MIR}}\right) - \frac{\delta T_s}{T_a}\frac{T_w^{MIR}}{\varepsilon_w^{MIR}} \right) = \underbrace{(\varepsilon_a - 1) + (2 - \varepsilon_a)\frac{T_w^{VIS} + A_w^{VIS} \cdot \left(1 + \frac{T_w^{MIR}}{\varepsilon_w^{MIR}}\right)}{2T_w^{MIR} + \varepsilon_w^{MIR}}}_{X_w} \quad (23)$$

$$4\frac{\delta T_s}{T_a} + \underbrace{\frac{h}{2\sigma T_a^3 (2T_w^{MIR} + \varepsilon_w^{MIR})}}_{\chi} \left( \frac{\delta T_w}{T_a} + \frac{\delta T_s}{T_a} \right) = \underbrace{(\varepsilon_a - 1) + (2 - \varepsilon_a)\frac{2T_w^{VIS} + A_w^{VIS}}{2T_w^{MIR} + \varepsilon_w^{MIR}}}_{X_s}$$

and can be solved analytically to give the following expressions:

$$\frac{\delta T_w}{T_a} = \frac{4X_w + \chi\left(X_w + X_s \frac{T_w^{MIR}}{\varepsilon_w^{MIR}}\right)}{(4+\chi)(4+\chi(2+3\frac{T_w^{MIR}}{\varepsilon_w^{MIR}})) + \chi^2 \frac{T_w^{MIR}}{\varepsilon_w^{MIR}}}$$

$$\frac{\delta T_s}{T_a} = \frac{4X_s + \chi\left(X_s(2+3\frac{T_w^{MIR}}{\varepsilon_w^{MIR}}) - X_w\right)}{(4+\chi)(4+\chi(2+3\frac{T_w^{MIR}}{\varepsilon_w^{MIR}})) + \chi^2 \frac{T_w^{MIR}}{\varepsilon_w^{MIR}}} \quad (24)$$

$$\frac{\delta T_r}{T_a} = \frac{\delta T_s + \delta T_w}{2T_a} = \frac{2X_w + 2X_s + \chi X_s(1 + 2\frac{T_w^{MIR}}{\varepsilon_w^{MIR}})}{(4+\chi)(4+\chi(2+3\frac{T_w^{MIR}}{\varepsilon_w^{MIR}})) + \chi^2 \frac{T_w^{MIR}}{\varepsilon_w^{MIR}}}$$

### 4.1 Analysis of MODEL 3

1) A particular solution of Model 2 (Eq. 22) for which convection terms do not contribute is $T_w = T_s = T_w$ when choosing $T_w^{VIS} = \gamma \cdot T_w^{MIR}$, $A_w^{VIS} = \gamma \cdot \varepsilon_w^{MIR}$, for any values of $T_w^{MIR}, \varepsilon_w^{MIR}$ and $h$.

2) We now study how addition of convection influences maximal allowed transmission of the visible light $T_w^{VIS}$ through the window. As a design condition we demand that either the window temperature or the wall temperature or the room temperature do not exceed the atmospheric one $T_{w,s,r} < T_a$. From (24), without making any assumption on the strength of a convection coefficient we find that the abovementioned design conditions define the following allowed regions in the $T_w^{VIS}$, $A_w^{VIS}$ parameter space (equations in blue in (25)):

**Window temperature:** $\delta T_w < 0 \Rightarrow 4X_w + \chi\left(X_w + X_s \dfrac{T_w^{MIR}}{\varepsilon_w^{MIR}}\right) < 0$

$\left(1 + \dfrac{\chi}{4}\left(1 + 2\dfrac{T_w^{MIR}}{\varepsilon_w^{MIR}}\right)\right) \cdot T_w^{VIS} + \left(1 + \dfrac{T_w^{MIR}}{\varepsilon_w^{MIR}} + \dfrac{\chi}{4}\left(1 + 2\dfrac{T_w^{MIR}}{\varepsilon_w^{MIR}}\right)\right) \cdot A_w^{VIS} < 2\gamma\left(T_w^{MIR} + \dfrac{\varepsilon_w^{MIR}}{2}\right)\left(1 + \dfrac{\chi}{4}\left(1 + \dfrac{T_w^{MIR}}{\varepsilon_w^{MIR}}\right)\right)$

$\max\left(T^{VIS}\right)\Big|_{A^{VIS}=0} = 2\gamma\left(T_w^{MIR} + \dfrac{\varepsilon_w^{MIR}}{2}\right) \cdot f_w(\chi) < 2\gamma\left(T_w^{MIR} + \dfrac{\varepsilon_w^{MIR}}{2}\right) \quad \forall \chi$

where $f_w(\chi) = \dfrac{1 + \dfrac{\chi}{4}\left(1 + \dfrac{T_w^{MIR}}{\varepsilon_w^{MIR}}\right)}{1 + \dfrac{\chi}{4}\left(1 + 2\dfrac{T_w^{MIR}}{\varepsilon_w^{MIR}}\right)} \rightarrow \begin{cases} 1, \chi \rightarrow 0 \text{ - weak convection} \\ \dfrac{1 + \dfrac{T_w^{MIR}}{\varepsilon_w^{MIR}}}{1 + 2\dfrac{T_w^{MIR}}{\varepsilon_w^{MIR}}} \in \left(\dfrac{1}{2}, 1\right), \chi \rightarrow \infty \text{ - strong convection} \end{cases}$

**Wall temperature:** $\delta T_s < 0 \Rightarrow 4X_s + \chi\left(X_s(2 + 3\dfrac{T_w^{MIR}}{\varepsilon_w^{MIR}}) - X_w\right) < 0$

$\left(2 + \dfrac{3\chi}{4}\left(1 + 2\dfrac{T_w^{MIR}}{\varepsilon_w^{MIR}}\right)\right) \cdot T_w^{VIS} + \left(1 + \dfrac{\chi}{4}\left(1 + 2\dfrac{T_w^{MIR}}{\varepsilon_w^{MIR}}\right)\right) \cdot A_w^{VIS} < 2\gamma\left(T^{MIR} + \dfrac{\varepsilon^{MIR}}{2}\right)\left(1 + \dfrac{\chi}{4}\left(1 + 3\dfrac{T_w^{MIR}}{\varepsilon_w^{MIR}}\right)\right)$

$\max\left(T^{VIS}\right)\Big|_{A^{VIS}=0} = \gamma\left(T_w^{MIR} + \dfrac{\varepsilon_w^{MIR}}{2}\right) \cdot f_s(\chi) < \gamma\left(T_w^{MIR} + \dfrac{\varepsilon_w^{MIR}}{2}\right) \quad \forall \chi$

where $f_s(\chi) = \dfrac{1 + \dfrac{\chi}{4}\left(1 + 3\dfrac{T_w^{MIR}}{\varepsilon_w^{MIR}}\right)}{1 + \dfrac{3\chi}{8}\left(1 + 2\dfrac{T_w^{MIR}}{\varepsilon_w^{MIR}}\right)} \rightarrow \begin{cases} 1, \chi \rightarrow 0 \text{ - weak convection} \\ \dfrac{\dfrac{2}{3} + 2\dfrac{T_w^{MIR}}{\varepsilon_w^{MIR}}}{1 + 2\dfrac{T_w^{MIR}}{\varepsilon_w^{MIR}}} \in \left(\dfrac{2}{3}, 1\right), \chi \rightarrow \infty \text{ - strong convection} \end{cases}$

**Room temperature:** $\delta T_r < 0 \Rightarrow 2X_w + 2X_s + \chi X_s(1 + 2\dfrac{T^{MIR}}{\varepsilon^{MIR}}) < 0$

$\left(3 + \chi\left(1 + 2\dfrac{T^{MIR}}{\varepsilon^{MIR}}\right)\right) \cdot T^{VIS} + \left(2 + \dfrac{T^{MIR}}{\varepsilon^{MIR}} + \dfrac{\chi}{2}\left(1 + 2\dfrac{T^{MIR}}{\varepsilon^{MIR}}\right)\right) \cdot A^{VIS} < 4\gamma\left(T^{MIR} + \dfrac{\varepsilon^{MIR}}{2}\right)\left(1 + \dfrac{\chi}{4}\left(1 + 2\dfrac{T^{MIR}}{\varepsilon^{MIR}}\right)\right)$

$\max\left(T^{VIS}\right)\Big|_{A^{VIS}=0} = \dfrac{4}{3}\gamma\left(T^{MIR} + \dfrac{\varepsilon^{MIR}}{2}\right) \cdot f_r(\chi) < \dfrac{4}{3}\gamma\left(T^{MIR} + \dfrac{\varepsilon^{MIR}}{2}\right) \quad \forall \chi \qquad (25)$

where $f_r(\chi) = \dfrac{1 + \dfrac{\chi}{4}\left(1 + 2\dfrac{T^{MIR}}{\varepsilon^{MIR}}\right)}{1 + \dfrac{\chi}{3}\left(1 + 2\dfrac{T^{MIR}}{\varepsilon^{MIR}}\right)} \rightarrow \begin{cases} 1, \chi \rightarrow 0 \text{ - weak convection} \\ \dfrac{3}{4}, \chi \rightarrow \infty \text{ - strong convection} \end{cases}$

Tradeoff between window transmission and absorption in the visible can be summarized geometrically in Fig. 8. As follows from inequalities presented in (25), regions below the corresponding lines represent the range of values of $T_w^{VIS}, A_w^{VIS}$ for which either the window temperature or the wall temperature or the room temperature are smaller than the atmospheric one. For practical purposes, we are mostly interested in the room temperature and the wall temperature (shaded regions in Fig. 8). Intersection of the corresponding two regions represent the range of values of $T_w^{VIS}, A_w^{VIS}$ for which both the wall temperature and the room temperature are smaller than the atmospheric one. Both the window, the wall and the room temperatures attain that of the atmospheric one when $T_w^{VIS} = \gamma \cdot T_w^{MIR}$, $A_w^{VIS} = \gamma \cdot \varepsilon_w^{MIR}$. As clear from the graph, when $T_w^{VIS} \leq \gamma \cdot T_w^{MIR}$ and $A_w^{VIS} \leq \gamma \cdot \varepsilon_w^{MIR}$, it is guaranteed that both the window and the room temperatures will be smaller than the atmospheric temperature. Also, from (25) we observe that the maximal allowed value of the transmitted light in the visible reduces somewhat when convection is present. The new value represents a fraction of the original value when no convection is present.

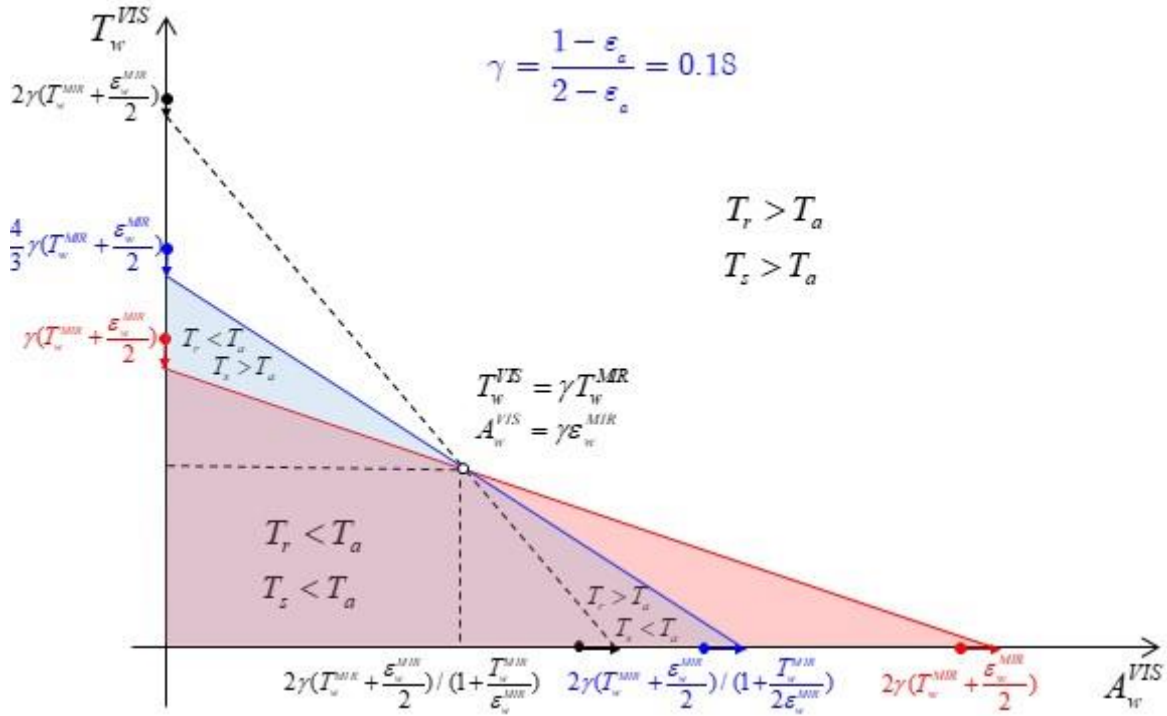

Fig. 8. Tradeoff between window transmission and absorption in the visible (with convection). A region of parameter space which is situated below the two lines define window operation regime for which both the wall and the room temperatures are smaller than that of the atmosphere. For completeness, a third dotted line is presented along which the window temperature is equal to that of the atmosphere.

3) We now study how addition of convection influences minimal achievable values of the window, wall and room temperatures in the problem of radiative cooling. Particularly, as follows from Model 2, radiative cooling is most efficient when window is almost perfectly reflective in the visible, and when window absorption loss in the visible is much smaller than that in the mid-IR, which can be summarized as $T_w^{VIS} \to 0, A_w^{VIS} \to 0, A_w^{VIS} \ll \varepsilon_w^{MIR}$. In the absence of convection, we then found that

$$\min\ T_w = \min\ T_s = \min\ T_r = \varepsilon_a^{1/4} \cdot T_a \approx 0.94 \cdot T_a\ ,\quad \text{or alternatively}$$

$$\min\left(\frac{\delta T_w}{T_a}\right) = \min\left(\frac{\delta T_s}{T_a}\right) = \min\left(\frac{\delta T_r}{T_a}\right) \approx \frac{\varepsilon_a - 1}{4} = 0.055.$$ When convection is present, from analysis of the linearized equations (24) it follows that for the radiative cooling to be most efficient, the same conditions as in the case without convection have to be satisfied $T_w^{VIS} \to 0, A_w^{VIS} \to 0, A_w^{VIS} \ll \varepsilon_w^{MIR}$. In this case, $X_s = X_w = \varepsilon_a - 1$ and from the following expressions for the minimal achievable window, wall, and room temperatures we conclude that the cooling efficiency degrades (minimal achievable temperatures grow):

$$\min\left(\frac{\delta T_w}{T_a}\right) = \frac{(\varepsilon_a - 1)}{4} \cdot \frac{1 + \frac{\chi}{4}\left(1 + \frac{T_w^{MIR}}{\varepsilon_w^{MIR}}\right)}{\left(1 + \frac{\chi}{4}\right)\left(1 + \frac{\chi}{4}\left(2 + 3\frac{T_w^{MIR}}{\varepsilon_w^{MIR}}\right)\right) + \left(\frac{\chi}{4}\right)^2 \frac{T_w^{MIR}}{\varepsilon_w^{MIR}}}$$

$$\min\left(\frac{\delta T_s}{T_a}\right) = \frac{(\varepsilon_a - 1)}{4} \cdot \frac{1 + \frac{\chi}{4}\left(1 + 3\frac{T_w^{MIR}}{\varepsilon_w^{MIR}}\right)}{\left(1 + \frac{\chi}{4}\right)\left(1 + \frac{\chi}{4}\left(2 + 3\frac{T_w^{MIR}}{\varepsilon_w^{MIR}}\right)\right) + \left(\frac{\chi}{4}\right)^2 \frac{T_w^{MIR}}{\varepsilon_w^{MIR}}} = \frac{(\varepsilon_a - 1)}{4} \cdot \begin{cases} 1, \chi \to 0, \text{weak convection} \\ \sim \frac{1}{\chi}, \chi \to \infty, \text{strong convection} \end{cases}$$

$$\min\left(\frac{\delta T_r}{T_a}\right) = \frac{(\varepsilon_a - 1)}{4} \cdot \frac{1 + \frac{\chi}{4}\left(1 + 2\frac{T_w^{MIR}}{\varepsilon_w^{MIR}}\right)}{\left(1 + \frac{\chi}{4}\right)\left(1 + \frac{\chi}{4}\left(2 + 3\frac{T_w^{MIR}}{\varepsilon_w^{MIR}}\right)\right) + \left(\frac{\chi}{4}\right)^2 \frac{T_w^{MIR}}{\varepsilon_w^{MIR}}}$$

(24)

## 5. Summary

We have considered several simple physical models of the radiative coolers and partially transparent radiative windows. We have then distinguished the design goals for these two structures and provided optimal choice for the material parameters to realise these goals. Out findings can be summarized as follows:

**Radiative cooling problem.** Radiative coolers are normally non-transparent, and the main question is how to design a cooler so that the temperature on its dark side is as low as possible compared to that of the atmosphere. This is achieved when window is almost perfectly reflective in the visible $R_w^{VIS} \to 1$ (alternatively $T_w^{VIS} \to 0, A_w^{VIS} \to 0$), and when window absorption loss in the visible is much smaller than that in the mid-IR $A_w^{VIS} \ll \varepsilon_w^{MIR}$. Then, in the absence of convection, and for any choice of the material optical parameters in the mid-IR ($T_w^{MIR}, \varepsilon_w^{MIR}$) $\min T_w = \min T_s = \varepsilon_a^{1/4} \cdot T_a \approx 0.94 \cdot T_a$. In the presence of convection, radiative cooling efficiency reduces potentially to zero when convection is strong.

**Partially transparent radiative window problem.** For the radiative windows, their surfaces are partially transparent, and the main question is rather about the maximal visible light transmission through the window at which the temperature on the window somber side does not exceed that of the atmosphere. This is achieved when window transmission in the mid-IR is almost perfect $T_w^{MIR} \to 1$ (alternatively $R_w^{MIR} \to 0, \varepsilon_w^{MIR} \to 0$), while window absorption in the visible is small $A_w^{VIS} \to 0$ and at the same time much smaller than that in the mid-IR $A_w^{VIS} \ll \varepsilon_w^{MIR}$. Then, in the absence of convection, the maximal allowed transmission through the window in the visible is $\gamma = \frac{1 - \varepsilon_a}{2 - \varepsilon_a} = 0.18$ while the wall, room and window temperatures are all smaller than the atmospheric one. If only requiring that the room and window temperatures are smaller than the atmospheric one, then the maximal transmission through the window in the visible can be increased to $\frac{4}{3}\gamma \approx 0.24$. In the presence of convection, maximal allowed window transmission in the visible reduces somewhat (at most as a certain multiplicative factor) even in the presence of strong convection.

**Effect of convection.** Convection contribution becomes important if the characteristic convection parameter is large $\frac{\chi}{4} = \frac{h}{8\sigma T_a^3 (2T_w^{MIR} + \varepsilon_w^{MIR})} > 1$, which is equivalent to $h > 6.46 \left[\frac{W}{K}\right] \cdot (2T_w^{MIR} + \varepsilon_w^{MIR})$. The value of the heat transfer coefficient depends strongly on the temperature differential, as well as the window size and orientation and is generally in the $1-10\left[\dfrac{W}{K}\right]$ range (see Appendix B for more details).

## 6. Acknowledgements

We would like to acknowledge financial support of the l'Institut de l'énergie Trottier and Canada Research Chairs of Prof. Skorobogatiy and Prof. Caloz for this project.

# APPENDIX A

**Solving radiation heat transfer equations for a single material plate in 1D.**

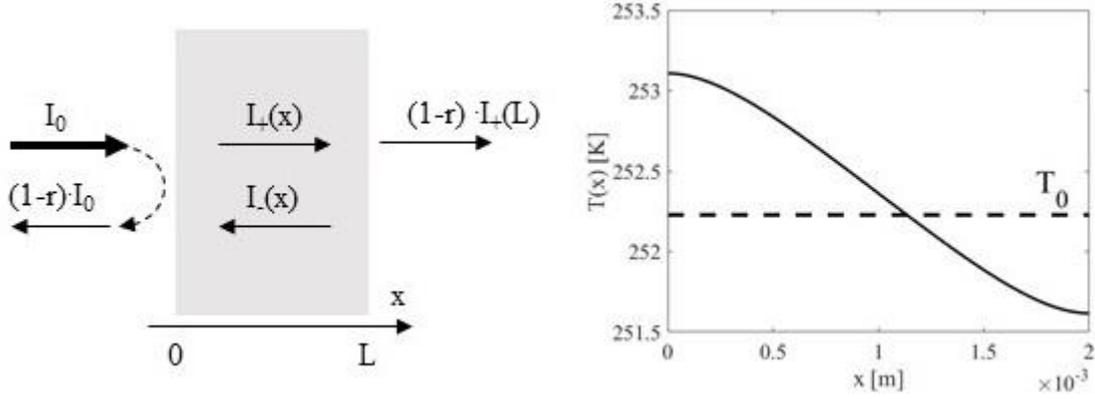

Fig A1. (Left) Schematic of an 1D radiative heat transfer problem. (Right) Typical temperature distribution across a thin plexiglass window.

Here we formulate radiative heat transfer equations in 1D for the case of a single plate irradiated with mid-IR radiation on the left (see Fig. A1.(a)). We then solve these equations in the limit of an almost-uniform temperature distribution inside a plate. The goal of this section is to demonstrate that simple expressions for the radiative fluxes used throughout the paper can be derived in 1D from the full radiative heat transfer formulation within the local thermodynamic equilibrium approximation.

Thus, following [47], we define $I_+(x)$ to be the average mid-IR radiative flux propagating in the positive direction, $I_-(x)$ to be the average mid-IR radiative flux propagating in the negative direction, and $T(x)$ to be the temperature distribution inside a plate. We consider that the plate is irradiated from the left by the mid-IR light with intensity $I_0$. Similar derivations can be performed when the plate is irradiated with the visible light, therefore we present one derivation. We then define the plate material absorption loss in the mid-IR as $\alpha$. The Fresnel reflection coefficient (by power) at the air/plate boundary is defined as $r = \left(\frac{n_m - n_a}{n_m + n_a}\right)^2$, where $n_m, n_a$ are the material and air refractive indices in the mid-IR. A single pass transmission through the plate is defined as $t = \exp(-\alpha L)$, where $L$ is the plate's thickness and $\alpha = 2\frac{\omega}{c}\text{Im}(n_m)$. The plate's material thermal conductivity is defined as $k$. Finally, we assume that the plate's medium is non-scattering.

Taking into the account material absorption of the mid-IR light as well as thermal mid-IR re-emission, for the two radiative heat fluxes we can write:

$$\frac{\partial I_+(x)}{\partial x} = -\alpha I_+(x) + \alpha \sigma T^4(x)$$
$$\frac{\partial I_-(x)}{\partial x} = \alpha I_-(x) - \alpha \sigma T^4(x)$$
(a.1)

Furthermore, from the energy conservation it follows that:

$$k\frac{\partial^2 T(x)}{\partial x^2} + \alpha(I_+(x) + I_-(x)) = 2\alpha \sigma T^4(x) \quad (a.2)$$

In the absence of heat conduction at the boundaries of a plate we assume thermally isolating boundary conditions:

$$\left.\frac{\partial T(x)}{\partial x}\right|_0 = \left.\frac{\partial T(x)}{\partial x}\right|_L = 0 \quad (a.3)$$

At the same time, incoming, forward and backward radiative fluxes are related through the optical boundary conditions at the interfaces expressed using Fresnel reflection coefficient $r$ as follows:

$$I_+(0) = (1-r)I_0 + rI_-(0)$$
$$I_-(L) = rI_+(L) \quad (a.4)$$

Equations (a.1-2) with boundary conditions (a.3-4) can be solved numerically. A typical temperature distribution across the plate is shown in Fig. A1.(b) for the case of a polycarbonate window $n_m = 1.5$, $r = 0.04$, $\alpha = 3200 m^{-1}$, $L = 2mm$, $t = 0.017$, $k = 0.2 \frac{W}{mK}$ and illumination intensity $I_0 = 459 \frac{W}{m^2}$ which corresponds to that of a black body at $300^0 C$. From the figure we see that the temperature at the plate's illuminated side is somewhat higher than the temperature at the plate's somber side, while the overall temperature variation inside of the plate is rather weak for this choice of the material's parameters. In fact, if we assume that the temperature inside of a plate is constant and equal to $T_0$, then equations (a.1) can be integrated exactly, thus resulting in purely exponential spatial dependence of the forward and backward propagating fluxes. Finally, integrating (a.2) over the plate thickness, and using boundary conditions (a.3-4) allows us to find the following expressions for the two outgoing fluxes (to the left and to the right of the plate):

$$I_0 = 2\sigma T_0^4$$

total energy flux in the $+\infty$ direction:
$$(1-r)I_+(L) = T_p \cdot I_0 + A_p \cdot \sigma T_0^4$$

total energy flux in the $-\infty$ direction:
$$rI_0 + (1-r)I_-(0) = R_p \cdot I_0 + A_p \cdot \sigma T_0^4$$

where the plate net absorption $A_p$, reflection $R_p$ and transmission $T_p$ coefficients (by power) are given by the following expressions:

$$R_p = r + \frac{(1-r)^2 t^2 r}{1-r^2 t^2} = r[1 + (1-r)^2 t^2 + (1-r)^2 r^2 t^4 + (1-r)^2 r^4 t^6]$$

$$T_p = t\frac{(1-r)^2}{1-r^2 t^2} = (1-r)^2 t(1 + r^2 t^2 + \cdots)$$

$$A_p = \varepsilon_p = \frac{(1-t)(1-r)}{1-rt} = 1 - R_p - T_p$$

Thus derived expressions for the plate reflection, absorption and transmission coefficients also correspond exactly to those that can be derived by assuming multiple ray passage through a thick absorbing optical medium and summing all the contribution of the partially transmitted and reflected rays (see Fig A2).

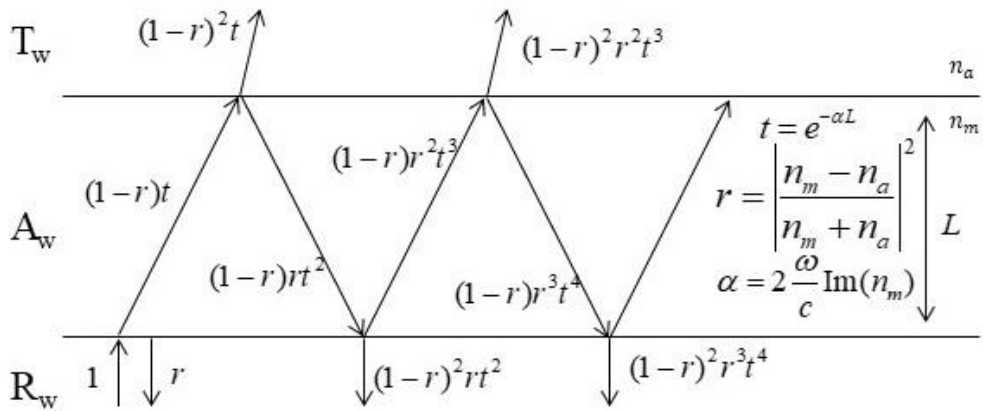

Fig A2. Transmission through and reflection from the optically thick plate (phase information is lost) can be described via addition of the energy fluxes of all the partially transmitted and reflected rays bouncing infinitely inside the plate.

## Appendix B

Models for the temperature dependent heat transfer coefficients at the planar gas/solid interfaces are well known and can be found for example in [47]. In what follows we present expressions that assume laminar flows along either horizontal or vertical interfaces. For both orientations a key parameter is a Rayleigh number [48] that can be expressed as follows:

$$Ra_L = \frac{g\beta}{\nu\alpha}(T_s - T_\infty)L^3$$

where $\beta = \frac{1}{T_f}$ is the expansion coefficient for ideal gases, $\nu$ is the kinematic viscosity, $\alpha$ is the thermal diffusivity and $L$ is the ratio of the plate surface area to its perimeters, $k$ is heat conductivity of air, $g = 9.8 \left[m/s^2\right]$ is the free fall acceleration. As the gas properties are temperature dependent, they are evaluated at the so-called film temperature $T_f = \frac{T_s + T_\infty}{2}$, which is the average of the surface $T_s$ and the surrounding bulk temperature $T_\infty$. The following are expressions for the various parameters of air:

$$\nu(T_f) = \mu(T_f)/\rho(T_f), \text{ where dynamic viscosity } \mu(T_f) = 2.55628 \times 10^{-7} \times T_f^{0.741} \left[Pa \cdot s\right]$$

$$k(T_f) = k_0 T_f^{0.8755} = 1.7254 \times 10^{-4} \times T_f^{0.8755} \left[W/(m \cdot K)\right]$$

$$\rho(T_f) = \frac{P_{atm}}{R \cdot T_f} = \frac{352.97}{T_f}\left[\frac{kg}{m^3}\right] \quad ; \quad C_p = \frac{7}{2}R \approx 1005 \frac{J}{kg \cdot K} \quad ; \quad \alpha(T_f) = \frac{k(T_f)}{\rho(T_f)C_p}$$

Then, for the Ryleigh number at the solid/air interface we find:

$$Ra = \frac{C_P \cdot \frac{1}{T_f} \cdot \left(\frac{P_{atm}}{R \cdot T_f}\right)^2 g}{k_0 \cdot \mu_0 \cdot T_f^{0.74} \cdot T_f^{0.88}} \approx \frac{g \cdot \frac{7}{2} P_{atm}^2}{R \cdot k_0 \cdot \mu_0} \times \frac{1}{T_f^{4.6174}} \approx \frac{2.7769 \times 10^{19}}{T_f^{4.6174}}(T_s - T_\infty)L^3$$

**Heat transfer coefficient for the horizontal plate (Laminar flow)**

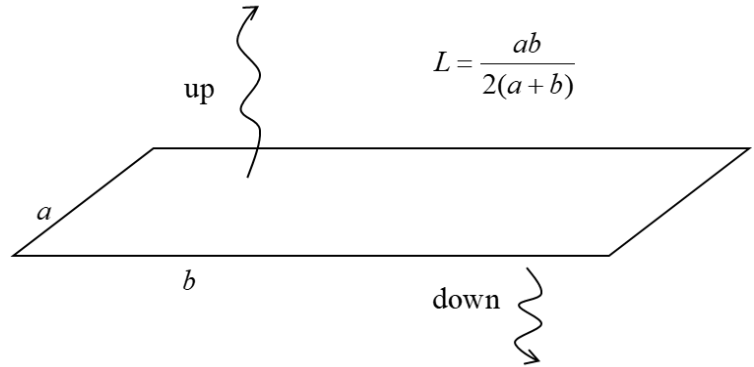

Hot surface facing up (or cold facing down),

$$h_{horisontal} = \frac{k}{L} \cdot 0.54 \cdot Ra_L^{\frac{1}{4}}, \qquad 10^5 < Ra_L < 2 \times 10^7$$

As hot air is moving up and cold air is moving down, convection heat transfer is more efficient for hot surfaces facing up or cold surfaces facing down.

Alternatively, for the hot surface facing down (or cold facing up)

$$h_{horisontal} = \frac{k}{L} \cdot 0.27 \cdot Ra_L^{\frac{1}{4}}$$

Thus, for the two cases we get the following expressions for the heat transfer coefficients:

$$h = \binom{2}{1} \cdot 3.33815 \cdot \left(\frac{T_s - T_\infty}{L \cdot T_f}\right)^{\frac{1}{4}} \cdot \frac{1}{T_f^{0.0289}}$$

**Heat transfer coefficient for the vertical plate (Laminar flow)**

$$h_{horisontal} = \frac{k}{L}(0.68 + \frac{0.67}{\left(1 + \left(\frac{0.492}{\Pr}\right)^{\frac{9}{16}}\right)^{\frac{4}{9}}} \cdot Ra_L^{\frac{1}{4}})$$

where the Pratt number for air is defined as:

$$\Pr = \frac{\nu}{\alpha} = \frac{\mu C_p}{k} = \frac{7}{2} R \frac{\mu_0 T_f^{0.74}}{k_0 T_f^{0.88}} \approx \frac{7}{2} \frac{R}{T_f^{0.1336}} \frac{\mu_0}{k_0} \approx \frac{1.4924}{T_f^{0.1336}}$$

Thus, leading to the following approximative expression for the heat transfer coefficient at the air/solid interface (inclined at an angle $\theta$ with respect to the vertical ($\theta \in (0, 60^\circ)$)):

$$h = \frac{1.1733 \times 10^{-4}}{L} T_f^{0.8755} + 6.4273 \left(\frac{T_s - T_\infty}{LT_f}\right)^{\frac{1}{4}} \frac{(\cos\theta)^{\frac{1}{4}}}{T_f^{0.0289}}$$

At $T_f \approx 300K$, the first term on the right hand in the expression is normally small (for example it equals to $0.017$ when $L = 1m$, and it equals to $0.17$ when $L = 0.1m$). The second term is normally much larger (it equals to $2.73$ when $L = 1m$ and $T_s - T_\infty = 1^oC$, and it equals to $4.86$ when $L = 0.1m$ and $T_s - T_\infty = 10^oC$).